\documentclass[aps,preprint,11pt]{revtex4}

\usepackage{amssymb,epsf}
\usepackage{epsfig}
\usepackage{epstopdf}
\usepackage{graphicx}
\usepackage{color}

\begin{document}

\title{\textbf{\Large Properties of defect modes in periodic lossy multilayer with negative index materials}}
\author{A. Aghajamali $^{1,2}$\footnote{email address:
alireza.aghajamali@fsriau.ac.ir} and M. Barati $^{2}$\footnote{email
address: barati@susc.ac.ir}}

\affiliation{\small $^1$ Young Researchers Club, Science and Research Branch, Islamic Azad University, Fars, Iran\\
$^2$ Department of Physics, Science and Research Branch, Islamic Azad University, Fars, Iran\\
}

\begin{abstract}
Transmission properties of one-dimensional lossy photonic crystals
composed of negative and positive refractive index layers with one
lossless defect layer at the center of the crystal are investigated
by the characteristic matrix method. The results show that as the
refractive index and thickness of the defect layer increase the
frequency of the defect mode decreases. In addition, it is shown
that the frequency of the defect mode is sensitive to the incidence
angle, polarization and physical properties of the defect layer but
it is insensitive to the small lattice loss factor. The height of
the defect mode is very sensitive to the loss factor, incidence
angle, polarization, refractive index and thickness of the defect
layer. It was also shown that the height and the width of the defect
mode are affected by the number of the lattice period and the loss
factor. The results can lead to designing new types of narrow filter
structures and other optical devices.
\end{abstract}\

\maketitle

\section{Introduction}

Photonic crystals (PCs) were first introduced theoretically by
Yablonovitch \cite{re1}., and experimentally by John \cite{re2}.
PCs, constructed with periodic structure of artificial dielectrics
or metallic materials, have attracted many researchers in the past
two decades for their unique electromagnetic properties and
scientific and engineering applications
\cite{re1,re2,re3,re4,re5,re6}. These crystals indicate a range of
forbidden frequencies, called photonic band gap, as a result of
Bragg scattering of the electromagnetic waves passing through such a
periodical structure \cite{re7,re8}. As the periodicity of the
structure is broken by introducing a layer with different optical
properties, a localized defect mode will appear inside the band gap.
Enormous potential applications of PCs with defect layers in
different areas, such as light emitting diodes, filters and
fabrication of lasers have made such structures are interesting
research topic in this field.

Negative index materials (NIMs) with simultaneously negative
permittivity and permeability, known as left-handed materials or
metamaterials, were introduced by Veselago \cite{re9} nearly forty
years ago. After the experimental realization of metamaterials by
Smith et al. \cite{re10}, such materials have received attention for
their very unusual electromagnetic properties
\cite{re11,re12,re13,re14,re15,re16,re17}. By possibility of
manufacturing PCs with metamaterials, called metamaterial photonic
crystals (MPCs) in recent years, a new research area was opened to
researchers where numerous interesting results have been reported so
far. Among the papers published in MPCs properties, there are a
number of reports in such structures with defect layers. The
properties of the defect modes in different one-dimensional (1D)
conventional PC and 1D MPC have been reported by several authors
\cite{re18,re19,re20,re21,re22,re23,re24,re25,re26,re27,re28,re29,re30,re31,re32,re33,re34}.

In this paper the characteristic matrix method is used to
investigate the effects of the lossless negative and positive defect
layers on the defect modes in 1D lossy MPC transmission spectra. The
paper is organized as follows: the MPC structure and theoretical
formulation (characteristic matrix method) are described in Section
2, the numerical results and discussions are presented in Section 3
and the paper is concluded in Section 4.

\section{MPC structure and characteristic matrix method \label{Bound}}

A schematic diagram of a one-dimensional MPC structure with defect
layer is shown in figure 1. The MPC structure is composed of
alternative NIM layers (layers \emph{A}) and PIM layers (layers
\emph{B}), where NIM is dispersive and dissipative. The thickness
and the refractive index of the layers are $d_A$ and $d_B$, $n_A$
and $n_B$, respectively. Layer \emph{C} is a defect layer and is
assumed to be a lossless and non dispersive material with thickness
and refractive index of $d_C$ and $n_C$, respectively. The
calculations are performed using the characteristic matrix method
\cite{re34}, which is the most effective technique to analyze the
transmission properties of PCs. The characteristic matrix of the
structure is given by:

$M[d]=(M_{A} M_{B})^{N/2} M_{C} (M_{B} M_{A})^{N/2}$

where $M_{A}$, $M_{B}$ and $M_{C}$ are the characteristic matrices
of layers \emph{A}, \emph{B}, and \emph{C}. The characteristic
matrix $M_{i}$ for TE waves at incidence angle $\theta_{0}$ in
vacuum is given by
\begin{equation}
M_i= \left[
\begin{array}{cc}
\cos \gamma_{i}  &  \frac{-i}{p_{i}} \ \sin \gamma_{i}\\
-i \ p_{i} \ \sin \gamma_{i}  &  \cos \gamma_{i}
\end{array}\right]
\end{equation}
where $\gamma_{i}=(\omega /c) \: n_{i} d_{i} \cos\theta_{i} $,
\emph{c} is speed of light in vacuum, $\theta_{i}$ is the angle of
refraction inside the layer \emph{i} with refractive index $n_i$ and
$p_{i}=\sqrt{\varepsilon_{i}/ \mu_{i}}\:\cos\theta_{i}$, where
$\cos\theta_{i}=\sqrt{1-(n_{0}^2\:\sin^2\theta_{0}/{n_{i}^2})}$. The
refractive index is given as $n_i=\pm \sqrt{\varepsilon_{i}\mu_{i}}$
\cite{re30,re35}, where the positive and negative signs are assigned
for the PIM and NIM layers, respectively.

The final characteristic matrix for an \emph{N} period structure is
given by:
\begin{equation}
[M(d)]^N= \prod^N_{i=1} M_i \equiv \left(
\begin{array}{cc}
m_{11}  &  m_{12}\\
m_{21}  &  m_{22}
\end{array}\right)
\end{equation}
where $m_{i,j}(i,j=1,2)$ are the matrix elements of $[M(d)]^N$. The
transmission coefficient of the multilayer is calculated by:
\begin{equation}
t=\frac{2\ p_{0}}{(m_{11}+m_{12}\ p_{s})\ p_{0}+(m_{21}+m_{22}\
p_{s})}
\end{equation}
where $p_{0}=n_{0}\ \cos\theta_{0}$ and $p_{s}=n_{s}\
\cos\theta_{s}$. The transmissivity of the multilayer is given by
\begin{equation}
T=\frac{p_{s}}{p_{0}} |t|^2 .
\end{equation}

The transmissivity of the multilayer for TM waves can be obtained by
changing $p_{i}=\sqrt{\mu_{i}/\varepsilon_{i}}\cos\theta_{i}$,
$p_{0}=\cos\theta_{0}/n_{0}$, and $p_{s}=\cos\theta_{s}/n_{s}$.

The permittivity and permeability of layer \emph{A} with negative
refracting index in the microwave region are complex and are defined
as \cite{re36},
\begin{equation}
\label{nineeq}
      \varepsilon_{A}(f) =1+\frac{5^2}{0.9^2-f^2-i\gamma f}+\frac{10^2}{11.5^2-f^2-i\gamma f}
\end{equation}
\begin{equation}
\label{teneq}
      \mu_{A}(f) =1+\frac{3^2}{0.902^2-f^2-i\gamma f}
\end{equation}
where \emph{f} and $\gamma$ are the frequency and damping frequency
in GHz, respectively. The behaviours of the real parts of the
permittivity and permeability of layer \emph{A}, $\varepsilon_{A} ^
{'}$ and $\mu_{A} ^ {'}$, versus frequency have been discussed in
our previous report \cite{re37}.

\section{Numerical results and discussion \label{Bound}}

The transmission spectrum of the MPC structure was calculated based
on the theoretical model described on the previous section. The
calculations are carried out in the region where $\varepsilon_{A} ^
{'}$ and $\mu_{A} ^ {'}$ are simultaneously negative, where a new
gap, called the zero-$\bar{n}$ gap \cite{re36}, will appear
\cite{re36,re37,re38,re39}. Layer \emph{B} is assumed to be the
vacuum layer with $\varepsilon_{B}=\mu_{B}=1$. The thickness of
layers \emph{A} and \emph{B} are chosen as $d_A=6$mm and $d_B=12$mm,
respectively. The defect layer \emph{C} is either NIM or PIM non
dispersive and non dissipative. The total number of the lattice
period is selected to be $N=16$ \cite{re37}.

The transmission spectrum for $d_C=24$mm with positive refractive
index of $n_C=0.75, 1.25, 1.75$ and $2.25$, for two different loss
factors ($\gamma = 0.2\times10^{-3}$ GHz and $\gamma =
8\times10^{-3}$ GHz) and for normal incidence are shown in figure 2.
As it is clearly seen, the rate of the transmittance and the height
of the defect modes are affected by the loss factor. The frequency
of the defect modes as a function of the refractive index of the
defect PIM and NIM layers for normal incidence and for two different
small lattice's loss factors are show in figure 3. As it is seen,
the frequency of the defect modes decreases as $n_C$ increases and
the modes are nearly insensitive to the small loss factors. In
addition, for the PIM defect layer (figure 3(a)) two defect modes
appear for some specified refractive indices. The behaviour is quite
different in the NIM defect layer (figure 3(b)) where no defect
modes are observed in some specific regions.

The effects of $n_C$ on the height of the defect modes for regions
I, II and III are shown in figure 3(a) for PIM defect layer and for
two different loss factors in figure 4. It is seen that the height
of the defect modes almost quadratically depends on the refractive
index. The minimum height decreases for the regions with large
$n_C$. Moreover, the height of the modes is very sensitive to the
loss factor and decreases as $\gamma$ increases. The transmission
spectrum for $n_C=1$ and $d_C=4, 16, 24$ and $32$mm and for two
different loss factors ($\gamma = 0.2\times10^{-3}$ GHz and $\gamma
= 8\times10^{-3}$ GHz) are presented in figure 5. As it is seen, the
frequency of the defect modes move toward the lower frequencies
(figure 5(a)) and the height of modes decrease as the thickness of
the defect layer increases (figure 5(b)). The behaviors of the
frequency and height of the defect modes versus $d_C$ for two
different $\gamma$ are shown in figures 6 and 7, respectively. As it
is seen, the frequency of the defect modes is nearly insensitive to
the loss factors but the height is very sensitive to $\gamma$ and
decreases as the loss factors increases.

In this section, the effects of the incidence angle on the defect
transmission spectra are investigated. The transmission spectra of
TE and TM polarized waves for $0^\circ$, $25^\circ$, $50^\circ$ and
$75^\circ$ incidence angles for $d_C=8$mm, $n_C=2.5$, and for two
different loss factors ($\gamma = 0.2\times10^{-3}$ GHz and $\gamma
= 8\times10^{-3}$ GHz) are shown in figures 8 and 9, respectively.

As it is observed from figure 8, and reported in \cite{re37,re39},
the width, the depth and the central frequency of the band gap for
TE waves increase as the incidence angle increases. But, for TM
waves, the gap disappears for the incidence angles greater than
$\approx50^\circ$. As $\theta_{0}$ increases the width and the depth
of gap decrease and the central frequency shifts toward the higher
frequencies (figure 9). The behaviour of the defect mode's frequency
for TE and TM waves as a function of the incidence angle for two
different loss factors are shown in figure 10. Here it is clearly
seen that the frequency of the defect mode is nearly insensitive to
the small loss factor and it disappears for TM waves for the
incidence angle greater than $\approx50^\circ$. The change of the
height of the defect mode for both polarizations as a function of
the incidence angle for two different loss factors is shown in
figure 11. It is seen that the height of the defect modes decreases
for TE polarized waves while it increases for TM polarized waves as
the incidence angle increases. It is also seen that the height of
the modes are very sensitive to the loss factor for both
polarizations.

In the last part the effects of the number of the lattice period,
\emph{N}, and the loss factor on the transmission properties of the
defect mode are studied. So far $N=16$ was assumed. The results are
presented in figures 12 and 13 for normal incidence angle. As it is
seen from figure 12(a) and as reported in \cite{re37}, the width of
the band gap for $N>16$ is almost independent of \emph{N}. However
the depth and the height of the defect mode reduce as \emph{N}
increases for $\gamma = 0.2\times10^{-3}$ GHz. It is interesting to
note that the frequency of the defect mode remains unchanged but the
width of the mode decreases as the number of the lattice period
increases (figure 12(b)). The transmission spectra of the structure
for four different loss factors and for $N=16$ are shown in figure
13. As it is seen from figure 13(a) the width and the depth of the
band gap and the frequency of the defect mode are nearly insensitive
to small loss factor but the transmission is affected by $\gamma$
\cite{re37}. Moreover, the height of the defect mode decreases as
the loss factor increases (figure 13(b)).

\section{Conclusion}

In this paper the transmission properties of one-dimensional lossy
photonic crystals with negative and positive refractive indices
layers and a lossless defect layer at the center of the crystal were
investigated by characteristic matrix method. It was shown that by
adding a lossless defect layer to a 1D lossy MPC, a localized defect
mode appears inside the band gap. The position of the defect mode
depends on the physical properties of the defect layer such as the
refractive index and thickness but it is independent of the loss
factor. The results show that as the refractive index and thickness
of the defect layer increases the frequency of the defect mode
decreases. In addition, it was shown that the frequency and the
height of the defect mode are sensitive to the incidence angle and
polarization. For a PIM defect layer with the refractive index
between $4.75$ to $6.25$, two defect modes appear in a specific
range of the transmittance spectrum, while for an NIM defect layer
the spectrum is quite different where no defect modes is observed in
a certain range of the refractive index ($n_C=-2$ to $-3.5$). The
absence of the defect mode in some specified region is not clear and
is the subject of our future studies. It was also seen that:
\emph{i}) by increasing the incidence angle the frequency of the
defect mode increases. \emph{ii}) by increasing the thickness of the
defect layer the frequency of the defect mode moves toward lower
frequencies. \emph{iii}) the height of the defect modes is very
sensitive to the loss factors. \emph{iv}) the depth of the band gap
increases and consequently the height and the width of the defect
mode decrease, so the mode becomes narrower for $N>16$. \emph{v})
the width and the depth of the band gap and the frequency of the
defect modes are nearly insensitive to small loss factors but the
height of the defect mode is very sensitive to small $\gamma$ and
decreases as the loss factor increases. Such properties are quite
useful in designing new types of optical devices in microwave
engineering.


\newpage
\thispagestyle{empty}
\begin{figure}[tbp]
\epsfxsize=7cm \centerline{\includegraphics [width=10cm] {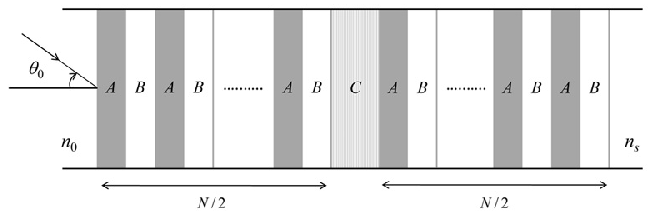}}
\caption{Schematic of 1D MPC structure with a defect layer. Layer
\emph{A} is NIM and layer \emph{B} is PIM materials. \emph{N} is the
number of lattice period and $\theta_{0}$ is the incidence angle.}
\end{figure}

\begin{figure}[tbp]
\epsfxsize=7cm \centerline{\includegraphics [width=15cm] {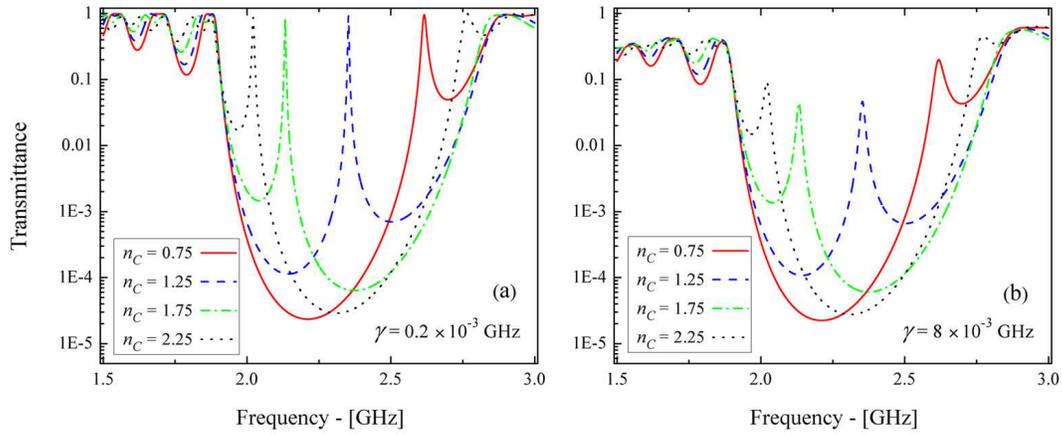}}
\caption{Transmission spectra of 1D MPC structure at normal
incidence for different positive refractive indices of the lossless
defect layer, with $d_C=24$mm and for two different lattice loss
factors (a) $\gamma = 0.2\times10^{-3}$ GHz (b)  $\gamma =
8\times10^{-3}$ GHz.}
\end{figure}

\begin{figure}[tbp]
\epsfxsize=7cm \centerline{\includegraphics [width=8cm] {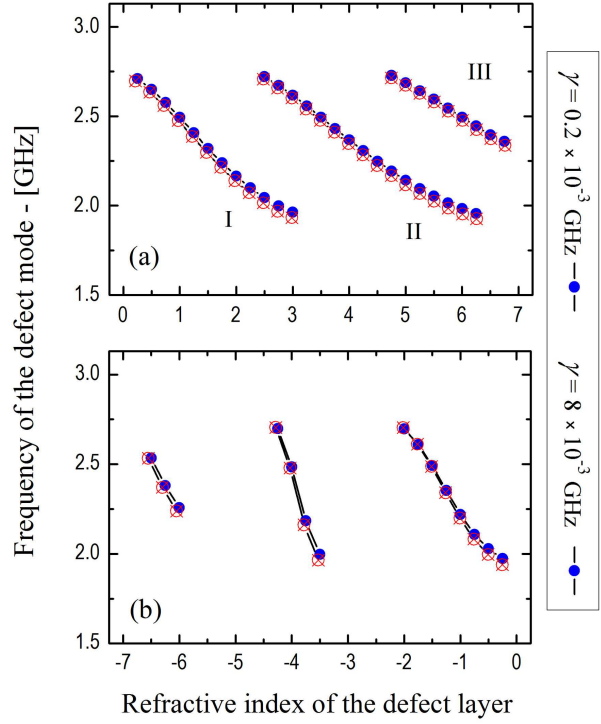}}
\caption{Dependence of the defect mode frequency on the (a) positive
and (b) negative refractive index of the lossless defect layer
($n_C$) for two different lattice loss factors ($\gamma =
0.2\times10^{-3}$ GHz and $\gamma = 8\times10^{-3}$ GHz).}
\end{figure}

\begin{figure}[tbp]
\epsfxsize=7cm \centerline{\includegraphics [width=15cm] {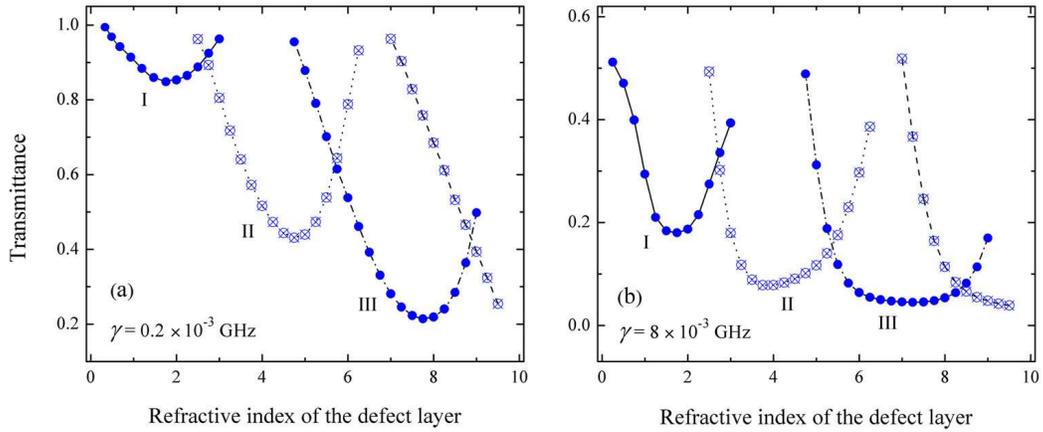}}
\caption{Height of the defect modes versus the positive refractive
index of the defect layer, with $d_C=24$mm and for two different
lattice loss factors (a) $\gamma = 0.2\times10^{-3}$ GHz (b) $\gamma
= 8\times10^{-3}$ GHz.}
\end{figure}

\begin{figure}[tbp]
\epsfxsize=7cm \centerline{\includegraphics [width=15cm] {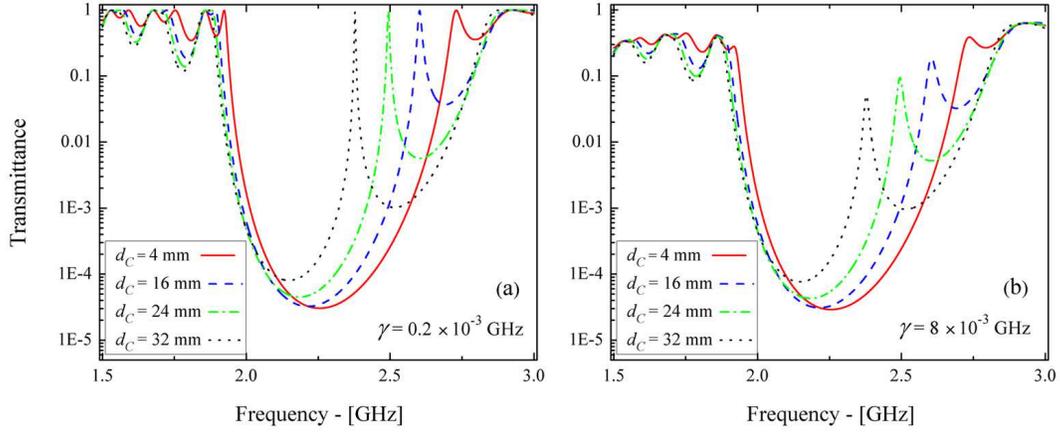}}
\caption{Transmission spectra of 1D MPC structure at normal
incidence, for different thicknesses of the lossless defect layer,
and for two different lattice's loss factors (a) $\gamma =
0.2\times10^{-3}$ GHz (b) $\gamma = 8\times10^{-3}$ GHz for with
$n_C=1$.}
\end{figure}

\begin{figure}[tbp]
\epsfxsize=7cm \centerline{\includegraphics [width=10cm] {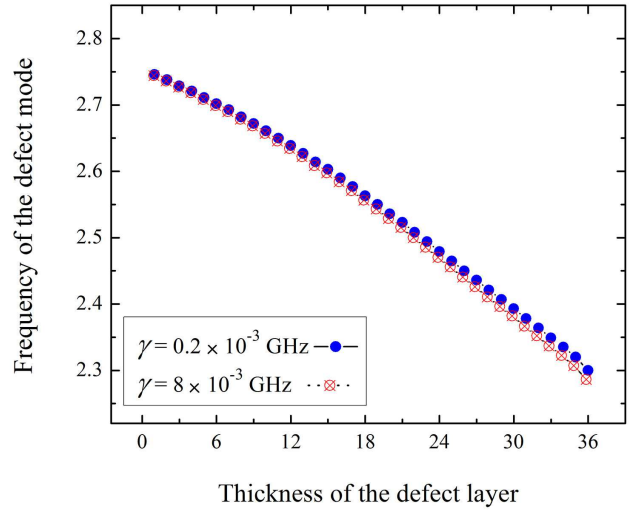}}
\caption{Dependence of the defect mode frequency on the thickness of
the defect layer for two different lattice loss factors for
$n_C=1$.}
\end{figure}

\begin{figure}[tbp]
\epsfxsize=7cm \centerline{\includegraphics [width=10cm] {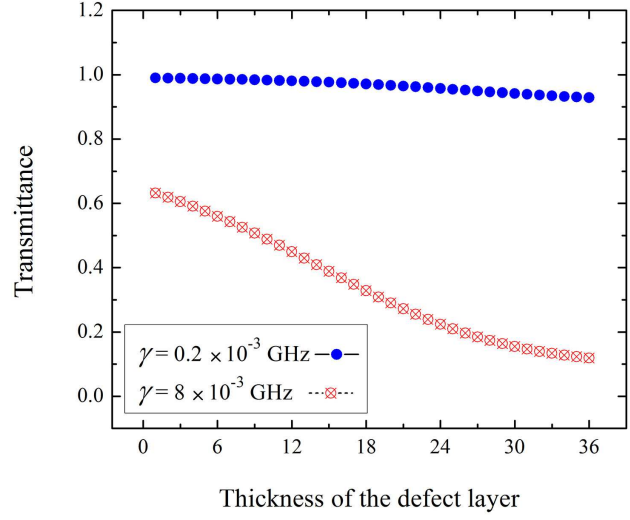}}
\caption{Height of the defect mode versus thickness of the defect
layer for two different lattice loss factors, for $n_C=1$.}
\end{figure}

\begin{figure}[tbp]
\epsfxsize=7cm \centerline{\includegraphics [width=15cm] {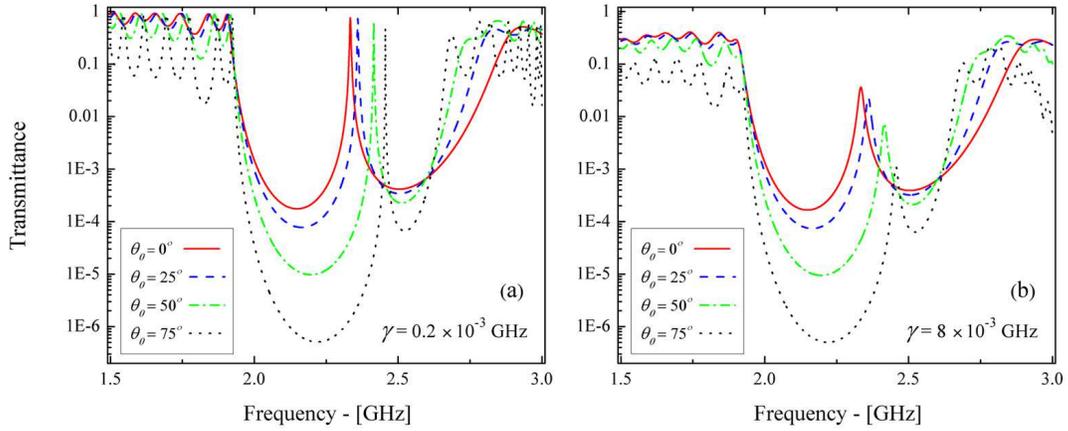}}
\caption{TE polarized wave transmission spectra of 1D MPC structure,
for different incidence angles with $n_C=2.5$ and $d_C=8$mm and for
two different lattice loss factors (a) $\gamma = 0.2\times10^{-3}$
GHz (b) $\gamma = 8\times10^{-3}$ GHz.}
\end{figure}

\begin{figure}[tbp]
\epsfxsize=7cm \centerline{\includegraphics [width=15cm] {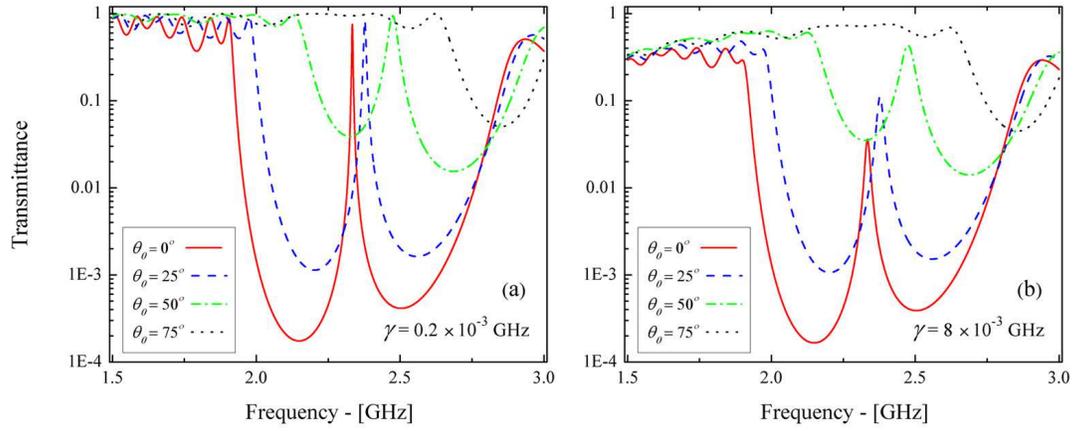}}
\caption{The same as figure 8 but for TM polarized wave.}
\end{figure}

\begin{figure}[tbp]
\epsfxsize=7cm \centerline{\includegraphics [width=10cm]
{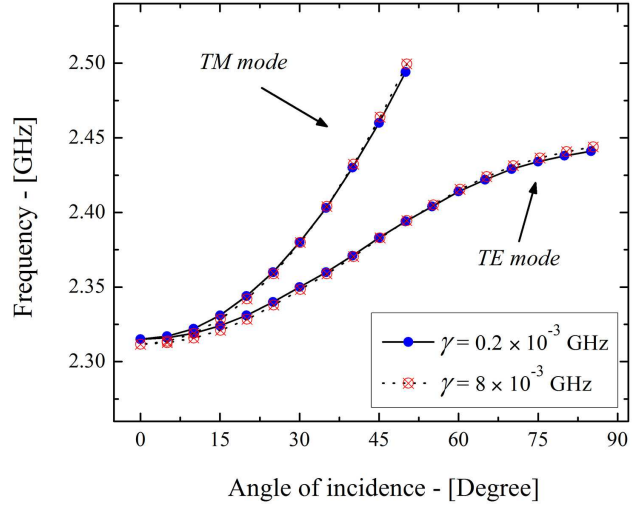}} \caption{Frequency of the defect mode as a function of
incidence angle for both polarizations and for two different lattice
loss factors.}
\end{figure}

\begin{figure}[tbp]
\epsfxsize=7cm \centerline{\includegraphics [width=10cm]
{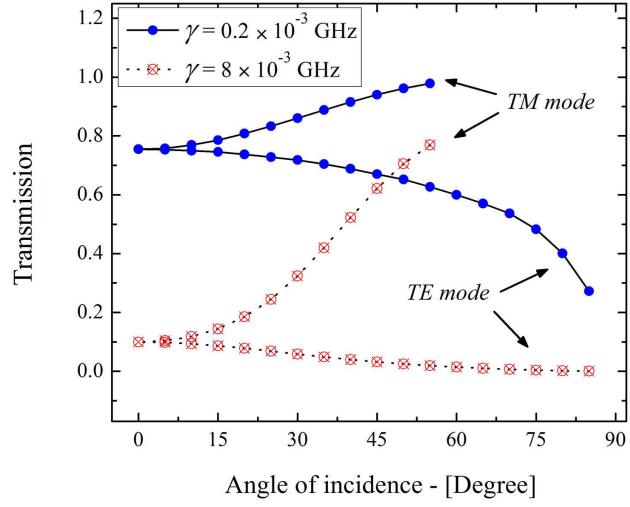}} \caption{Height of defect mode as a function of
incidence angle for both polarizations and for two different lattice
loss factors.}
\end{figure}

\begin{figure}[tbp]
\epsfxsize=7cm \centerline{\includegraphics [width=15cm]
{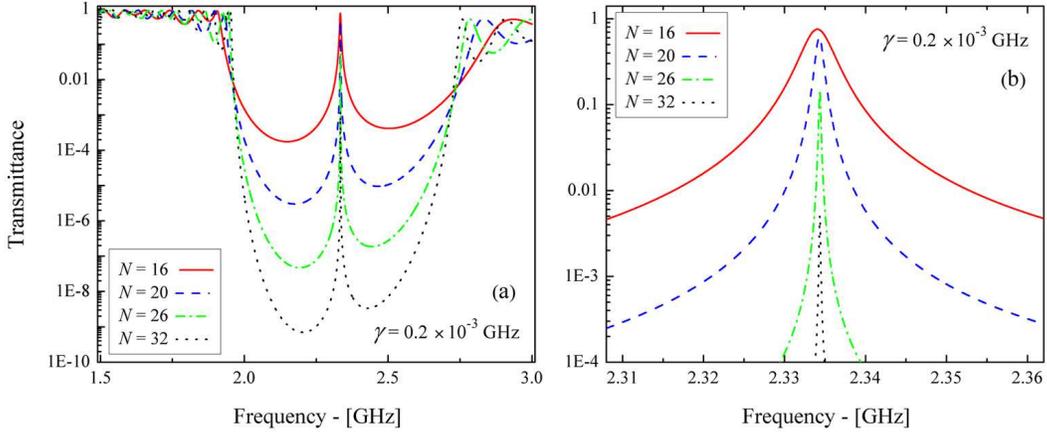}} \caption{Transmission spectra of 1D MPC structure at
normal incidence, for different number of the lattice period and for
$\gamma = 0.2\times10^{-3}$ GHz.}
\end{figure}

\begin{figure}[tbp]
\epsfxsize=7cm \centerline{\includegraphics [width=15cm]
{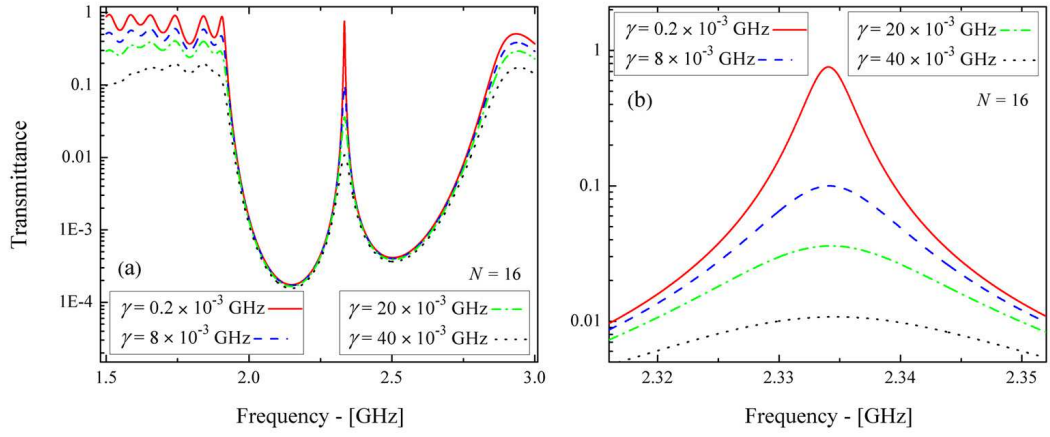}} \caption{Transmission spectra of 1D MPC structure at
normal incidence, for different lattice loss factors for $N=16$.}
\end{figure}

\end{document}